\documentclass[12pt,preprint]{aastex}

\newcommand{\be}{\begin{eqnarray}}
\newcommand{\ee}{\end{eqnarray}}
\newcommand{\br}{\mathbf{r}}

\newcommand{\bfv}{\mathbf{v}}

\newcommand{\bF}{\mathbf{F}}
\newcommand{\smass}{{M_{\odot}}}

\shorttitle{Triple Blackholes I}
\shortauthors{Iwasawa, et al. }

\begin{document}

\title{Evolution of Massive Blackhole Triples I --- Equal-mass
binary-single systems}

\author{Masaki Iwasawa \altaffilmark{1} , Yoko Funato\altaffilmark{1}
and Junichiro Makino\altaffilmark{2}
}
\email{iwasawa@margaux.astron.s.u-tokyo.ac.jp}

\altaffiltext{1}{Department of General System Studies, University of Tokyo,
3-8-1 Komaba, Meguro-ku, Tokyo 153-8902, Japan}
\altaffiltext{2}{Department of Astronomy, University of Tokyo,
7-3-1 Hongo, Bunkyo-ku, Tokyo 113-0033, Japan.}

\begin{abstract}

We present the result of $N$-body simulations of dynamical evolution
of  triple massive blackhole (BH) systems in galactic nuclei.  We
found that in most cases two of the three BHs merge through
gravitational wave (GW) radiation in the timescale much shorter than the
Hubble time, before ejecting one BH through a slingshot. In order for a
binary BH to merge before ejecting out the third one, it has to become
highly eccentric since the gravitational wave timescale would be much
longer than the Hubble time unless the eccentricity is very high. 
We found that two mechanisms drive the increase of the eccentricity of the
binary. One is the strong binary-single BH interaction resulting in
the thermalization of the eccentricity. The second is the Kozai mechanism
which drives the cyclic change of the inclination and eccentricity of
the inner binary of a stable hierarchical triple system. Our result
implies that many of supermassive blackholes are binaries.

\end{abstract}


\section{Introduction}

When two galaxies merge, the central BHs sink toward the center
of the merger remnant and form a binary system
(Begelman, Blandford, \& Rees 1980, henceforth BBR).
Whether or not this binary BH can merge in the
timescale shorter than the Hubble time is an important question for
the understanding of the evolution and growth of BHs and galactic
nuclei.

BBR predicted that the hardening of the BH binary through the stellar
dynamical interaction with field stars (FSs) would slow down after the loss
cone is depleted. After the depletion, the growth will be driven by
the refilling of the loss cone by two-body relaxation, and a simple
theoretical estimate results in the merger timescale orders of
magnitude larger than the Hubble time.

Whether or not this slowing down of evolution actually occur has been
studied with $N$-body simulations \citep{QH97,MilosavljevicMerritt2003},
Until very recently, however, there has 
been serious discrepancy between theory and simulation results, and
also among simulation results. Theory predicts that the timescale
must be proportional to the number of particles used to express the
parent galaxy, since the relaxation timescale is (neglecting the weak
$\log N$ term) proportional to the number of particles in the system
$N$. The results of $N$-body simulations range from no
dependence on $N$ \citep{Chatterjeeetal2003,MilosavljevicMerritt2001}
to $N^{1/3}$ \citep{Makino1997}.

\citet{MF04} performed $N$-body simulations with much larger number of
particles (up to $10^6$) and for longer time compared to previous
works, and concluded that the $N$-dependence of the evolution
timescale of BH binary is consistent with the theoretical
prediction. \citet{berczik2005} obtained similar result, by performing
even longer simulations for somewhat smaller $N$ (up to 256k).

Thus, as far as the stellar dynamical effects are concerned, BH
binary, once depleted the loss cone, would not merge within the Hubble
time, unless there remain lots of gas in the central region of
merger remnants.

If BHs in galaxies, in particular giant ellipticals, are mostly
binaries, growth of the BH through merging might be difficult. If one
galaxy with binary BH and the other with single BH merge, the central
BHs form a triple system, which is generally unstable. The most likely
outcome would be the ejection of at least one BH, 
preventing the growth of the central BH binary.
In fact, triple interaction of BHs was first considered
as an mechanism to eject BHs from the center of a
galaxy\citep{SaslawValtonenAarseth1974}. The timescale for the merging
through GW radiation is given by

\begin{equation}
\label{eqn:tgr}
t_{gr}=2\times 10^{15}\left(\frac{10^8M_{\odot}}{m_{B}}\right)^3
\left(\frac{a}{1\rm{pc}}\right)^4g(e) ,
\end{equation}
\begin{equation}
g(e)=\frac{(1-e^2)^{7/2}}{1+(73/24)e^2+(37/96)e^4} ,
\end{equation}
where $m_B$ is the mass of BHs (assuming equal masses), $a$ and $e$
are the semi-major axis and eccentricity of the binary. In the case of
a BH with $10^8 M_{\odot}$, the orbital separation must become less
than 0.1 pc in order for the merging timescale to become shorter than
the Hubble time, if we assume the average thermalized value of
$e=0.7$.

Since the gravitational wave timescale depends  on the
eccentricity very strongly, if some mechanism drives the binary BH to high
eccentricity the merging timescale might become short. \citet{ME94}
argued that the maximum eccentricity a binary BH would experience is
much higher than the average value, since a binary BH must undergo
around 10 strong single-binary scattering events before ejecting out
a single BH. They concluded, using simple analytic arguments, that
the binary BH in a triple BH system would merge, for the case of
typical BH of $10^8M_{\odot}$ in a typical elliptical galaxy within
the Hubble time. Another
possibility is the change of eccentricity through Kozai
mechanism\citep{kozaimec1}. 

In this paper, we investigate the evolution of triple BH systems by
means of direct $N$-body simulations.
Our initial model consists of an $N$-body realization of a spherical galaxy
and  three BH particles. We started with hierarchical configuration in
which two BH particles initially form a binary. The third BH particle
is placed at some distance to the binary. We changed the initial
orbit of the third particle, to see its effect on the evolution of the
binary. We incorporated the effect of GW radiation on
the relative orbits of BH particles.

We found that if the initial orbits of three BHs are coplanar and if
the total angular momentum is small, the binary BH hardens through
repeated binary-single BH interactions. The interaction
stops either when one BH is ejected out of the galaxy or when the BH
binary merges through GW radiation. Which of the two
dominates depends on the depth of the potential well of the parent
galaxy. Merging is more likely for deeper potential. For typical giant
ellipticals, potential is deep enough that merging dominates.

When the orbits are not coplanar, hierarchical triples are  formed
and the inner binary evolves  through Kozai
mechanism, in a fair fraction of cases. Kozai mechanism leads the
large-amplitude oscillation of eccentricity.  Thus, merging is more likely
in three-dimensional configuration than in coplanar systems.

The structure of this paper is as follows.  In section 2, we describe
the initial models
and method of our numerical simulation. In section 3, we
show the result of the cases in which three black holes are initially
in coplanar orbits. In section 4, we show the result of more
general cases of non-coplanar initial conditions.  We summarize our
result of numerical simulation in section 5.

\section{Initial Models and Numerical Methods}

\subsection{Models}

\begin{figure}
\plotone{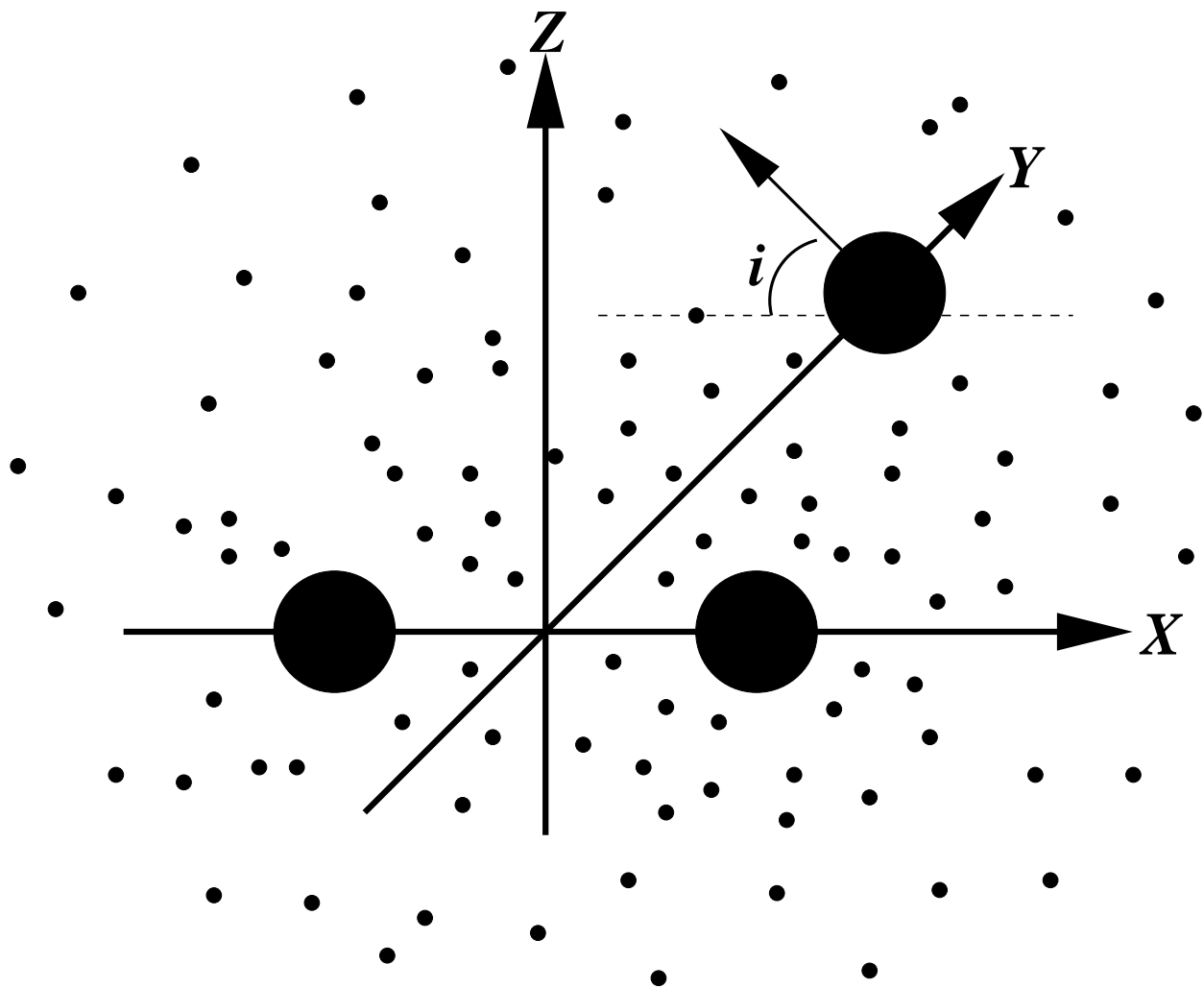}
\caption{ Initial setup of BH particles in  the parent galaxy is shown
schematically.
Small dots and large filled circles denote field particles and BH particles,
respectively. The arrow shows the initial velocity of the 3rd
blackhole. $i$ is the inclination. 
\label{fig:1}}
\end{figure}

The initial setup is schematically shown in figure \ref{fig:1}.
We placed three BH particles in $N$-body models of a spherical
galaxy. 
The quantitative properties of models and initial conditions of our
simulation is summarized in table 1.

For the galaxy model, we used a King model with $W_{0}=7$, where
$W_{0}$ is the nondimensional central potential of King models
\citep{kin66, GD}. We performed several runs changing the number of
particles in a galaxy from 16K to 128K.
We adopted the the standard
$N$-body units \citep{Hg}, in which  $M_{{\mathrm gal}}=1.0, E_{{\mathrm
gal}}=-0.25, G=1$. Here, 
$M_{{\mathrm gal}}$ and  $E_{{\mathrm gal}}$ are the total mass and
total binding energy of the galaxy and  $G$ is the gravitational constant.
The relation to the physical units will be discussed in
section 2.5.

We placed three blackholes in the parent galaxy model. All blackhole
particles have the same mass of 0.01 in the $N$-body units.
Two blackholes are initially placed 
as a binary with eccentricity
$e\sim 0.7$ in the
center of the galaxy model. 
The orbital plane of this binary is 
the $x-y$ plane.  the third blackhole is placed at position (1,0,0).

We have changed the initial velocity of the third particles in several
different ways. In the simplest case, it is placed at rest. In one
series, we gave initial velocity along $y$ axis, and in another series
along $z$ and in the third one in $y-z$ plane.
Tables \ref{tbl2} and \ref{tbl3} list all runs we performed.
We scaled the positions and velocities of stars so that the total energy of the
system, including that of black hole particles, becomes $-1/4$.  

\begin{deluxetable}{lcc}
\tablecaption{Model Parameters \label{tbl1}}
\tablewidth{0pt}
\tablehead{\colhead{Parameter} & \colhead{Symbol}   & \colhead{Value}}
\startdata
Mass of galaxy &$M_{gal}$ &$1.0$ \\
Mass of FS &$m_f$ &$1/16384-1/131072$  \\
Mass of BH &$m_B$ &$0.01$  \\
Number of FS &$N_{f}$ &$16-128k$ \\
Number of BH &$N_{B}$ &$3$ \\
Gravitational constant &$G$ &$1$ \\ 
Total energy &$E_{gal}$ &$-0.25$ \\
Softening between BH and BH &$\epsilon_{BB}$ &$0.0$ \\
Softening between BH and FS &$\epsilon_{Bf}$ &$10^{-6}$ \\
Softening between FS and FS &$\epsilon_{ff}$ &$0.01-0.001$ \\
\enddata
\end{deluxetable}

\subsection{Equations of Motion}

The equation of motion for field particles is given by
\begin{equation}
\frac{d^2{\br_{f,i}}} {dt^2}  =  {\bF_{ff,i}} + {\bF_{fB,i}} ,
\end{equation}
where ${\bF_{ff,i}}$ and ${\bF_{fB,i}}$ are acceleration due to field
stars and BH particles respectively. They are given by

\begin{equation}
{\bF_{ff,i}}   =  \sum_{j}^{N_f} \frac{{-m_{f,j}
 \br_{ff,ij}}}{\left(r^2_{ff,ij}+\epsilon_{ff}^2\right)^{3/2}},
\end{equation}
\begin{equation}
{\bF_{fB,i}}   =  \sum_{j}^{N_B} \frac{{-m_{B,j}
 \br_{fB,ij}}}{\left(r^2_{fB,ij}+\epsilon_{fB}^2\right)^{3/2}}, 
\end{equation}
\begin{equation}
{\br_{ff,ij}}  =  {\br_{f,i}} -  {\br_{f,j}}, 
\end{equation}
\begin{equation}
{\br_{fB,ij}}  =  {\br_{f,i}} -  {\br_{B,j}}.
\label{eqn:m1}
\end{equation}
Here $m_{f,i}$ and  ${\br_{f,i}}$
are the mass and the position of field star with index $i$, and
 $m_{B,i}$, ${\br_{B,i}}$ are those of black hole particle $i$.
We used softening parameters $\epsilon_{ff}$  and $\epsilon_{fB}$ for
field star-field star interactions and field star-black hole
interactions. Their values are  $\epsilon_{ff}=0.01$ and
$\epsilon_{fB}=10^{-6}$, respectively.

The equation of motion for BH particles is
\begin{equation}
\frac{d^2{\br_{B,i}}} {dt^2}  =  {\bF_{Bf,i}} + {\bF_{BB,i}},
\end{equation}
where ${\bF_{Bf,i}}$ and ${\bF_{BB,i}}$ are acceleration due to field
stars and BH particles, respectively. They are given by
\begin{eqnarray}
&&{\bF_{Bf,i}}   =  \sum_{j}^{N_f} \frac{{-m_{f,j} \br_{Bf,ij}}}
{\left(r^2_{Bf,ij}+\epsilon_{fB}^2\right)^{3/2}}, \\
&&{\bF_{BB,i}}   =  \sum_{j}^{N_B} 
\frac{{-m_{B,j} \br_{BB,ij}}}
{\left(r^2_{BB,ij}+
\epsilon_{BB}^2\right)^{3/2}}
+ \sum_{j}^{N_B} \bF_{GW,ij} .
\label{eqn:m2}
\end{eqnarray}
We set the softening between BH particles as $\epsilon_{BB}=0$.
The second term,
$\bF_{GW}$, is the quadrupole approximation for the orbital change due
to GW radiation \citep{Damour87}, which is  expressed as
\begin{eqnarray}
\label{eqn:gw}
\bF_{GW,ij} &=& \frac{4G^2m_{B,i}m_{B,j}}{5r^3c^5} 
\Biggl[\Biggl(
 - v^2 + \frac{2Gm_{B,i}}{r} -
 \frac{8Gm_{B,j}}{r}\Biggl)
{\bfv} , \nonumber \\
&& + \frac{{\br} \cdot {\bfv}}{r^2}
\Biggl( 3 v^2 - \frac{6Gm_{B,i}}{r}
+ \frac{52Gm_{B,j}}{3r} \Biggl) {\br}\Biggl] ,
\end{eqnarray}
\begin{equation}
{\br} = {\br}_{BB,ij} = {\br_{B,i}}-{\br_{B,j}} ,
\end{equation}
\begin{equation}
{\bfv} = {\bfv}_{BB,ij} = {\bfv_{B,i}}-{\bfv_{B,j}} , 
\end{equation}
where $\bfv_{B,i}$ is the velocity of black hole $i$.

\subsection{Numerical Method}

We integrated the equations of motions using the 4th order Hermite
scheme \citep{her} with individual variable time step. In order to
apply the Hermite scheme, we need the time derivative of the
acceleration. For the gravitational wave term (Eq. \ref{eqn:gw}), we
used the following form as the time derivative:
\begin{eqnarray}
\label{eqn:jerkgw}
\frac{d\bF_{GW,ij}}{dt}&=&\frac{4G^2m_{B,i}m_{B,j}}{5r^3c^5} 
 \Biggl[ \Biggl( 
- 2 \left( {\bfv} \cdot {\bf a} \right)
+ \frac{3 \left( {\br}
 \cdot {\bfv} \right) v^2}{r^2}
\Biggl) {\bfv}
- v^2 {\bf a}  \nonumber \\
&& + \frac{G}{r} (2m_{B,i} - 8m_{B,j})
\left( {\bf a} 
- \frac{4\left( {\br}\cdot {\bfv}\right)}{r^2}{\bfv}
 \right) \nonumber \\
&&+ \frac{3}{r^2}
\Biggl( \Big( \big( 
v^2+{\br}\cdot {\bf a} \big) v^2
+ 2\big( {\bfv}\cdot {\bf a}\big) \left( {\br}\cdot {\bfv} \right)
- \frac{5\big( {\br}\cdot {\bfv} \big)^2
v^2}{r^2} 
\Big) {\br}
+ v^2 \left({\br}\cdot {\bfv}\right)
{\bfv}\Biggl) \nonumber \\
&& + \frac{1}{r^3}\left( \frac{52}{3}Gm_{B,j} - 6Gm_{B,i}\right) \nonumber \\
&& \times \Biggl( \Big( v^2 + \left({\br} \cdot 
{\bf a}\right)\Big)
{\br} + \left( {\br}\cdot {\bfv}\right) {\bfv} 
 - \frac{6\left({\br}\cdot{\bfv}\right)^2}{r^2}{\br}
\Biggl)
\Biggl],
\end{eqnarray}
\begin{equation}
{\bf a} = {\bf a}_{BB,ij} = {\bf a}_{B,i}-{\bf a}_{B,j}, 
\end{equation}
where $\bf{a}_{B,i}$ is $i$th black hole acceleration.

In order to calculate the acceleration due to field particles, we used
GRAPE-6\citep{grape6}, the special purpose computer for the
gravitational $N$-body problem. Forces from BH particles, both
Newtonian and gravitational wave terms, are calculated on the host
computer.

In all runs, the energy (corrected for the loss of energy through
GW radiation)is conserved to 0.1\%.

\subsection{Correspondent Physical Scales}

We interpret the mass unit in our simulation as $10^{10} \smass$.
Thus, the mass of BH particles is $10^{8} \smass$. The total mass of
$10^{10} \smass$ corresponds to the mass of the nucleus of a galaxy
containing a BH. In other words, the galaxy model we use is not really
the model of the entire galaxy, but only that of the central
region. The reason we limit our model to the central region is simply
to reduce the number of particles $N$, in other words, to reduce the
calculation cost.  This modification means that
the potential depth of our model galaxy is somewhat less than
that of a real galaxy with the same velocity dispersion. To see the effect
of this difference, we have changed the velocity dispersion of the
galaxy and investigated the effect. In the standard runs, we set the
velocity dispersion of the initial King model as 300 km/s. In other
words, we set the light velocity $c$ in the $N$-body unit to 706. We
tried two different values of velocity dispersion, 600 km/s and 150
km/s, or the value of the light velocity 353 and 1412,
respectively. Note that by changing the value of $c$ actually we change the
strength of the GW term in equation (\ref{eqn:gw}) and thus change
the evolution of the BH particles.

\section{Result I --- Simplest configuration }

\label{sect:planar}

In this section, we describe the results of runs in which the third BH
is initially at rest, to illustrate the typical behavior and the
dependence of the results on the artificial parameters of numerical
experiments such as the softening and the number of particles. We
discuss the results of more general configurations in section
\ref{sect:general}. 

\begin{figure}
{\plotone{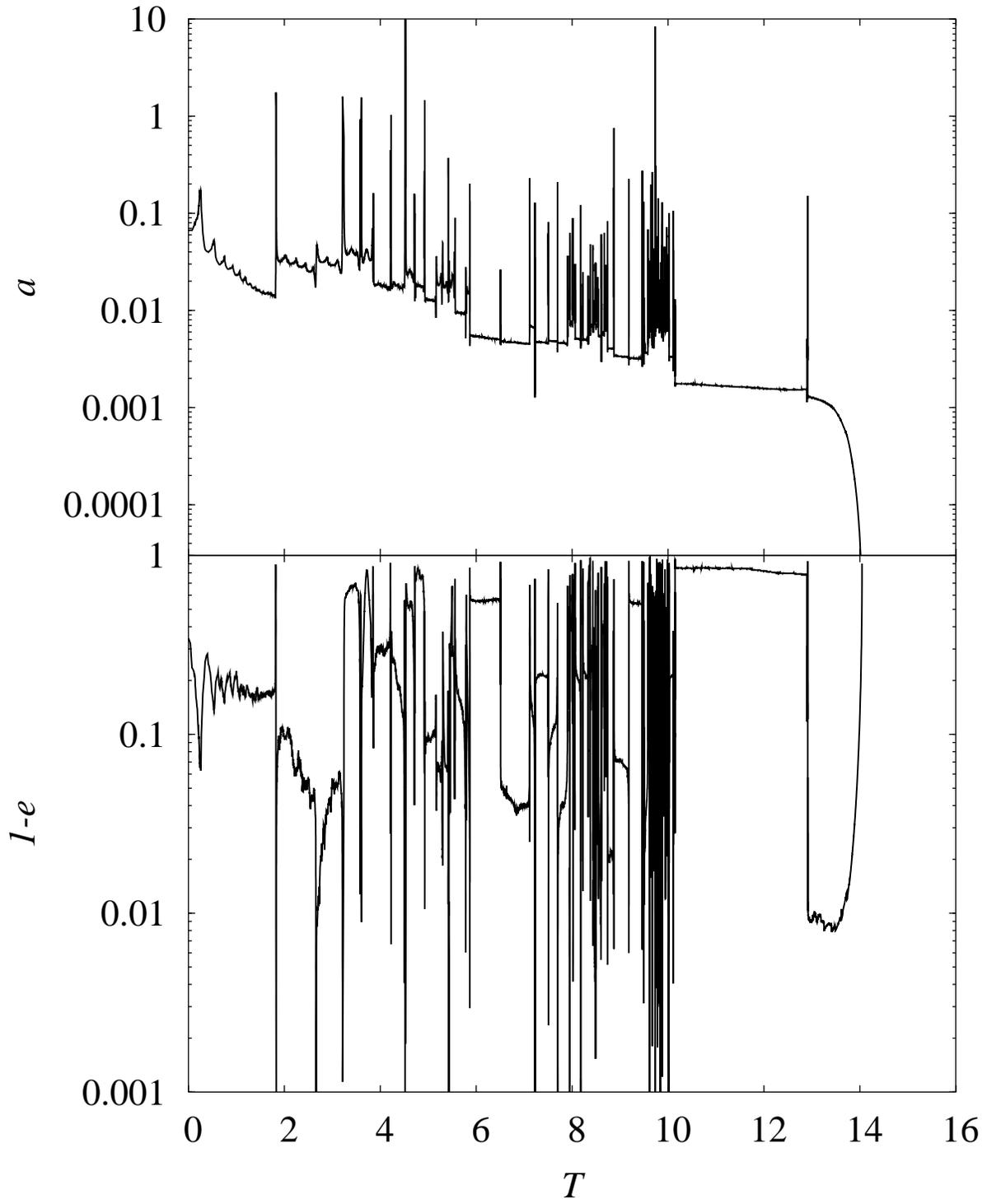}}
\caption{Evolution of semi-major axis $a$ (top panel) and
eccentricity $e$ (bottom panel, $1-e$ is shown) for run 64kW7A.
\label{fig2}}

\end{figure}

Figure \ref{fig2} shows the evolution of the semi-major axis $a$ and
the eccentricity $e$ of the BH binary 
in one run of model 64kW7 (hereafter called run 64kW7A). In this and all other
figures, the semi-major axis and eccentricity are defined as those of
the most strongly bound pair of BH particles.
Initially (for time
$T<1.8$), $a$ shows smooth decrease. This is due to the dynamical
friction from field stars. The jumps in $a$ and $e$ at $T\sim 1.8$ is
the result of first strong encounter between the binary and the third
body. In this case, the binary becomes wider and more eccentric. The
three BH particles experienced quite complex series of interactions
until $T\sim 10$. At this point, the third BH was ejected out, and it
took some time before it came back and interacted strongly with the binary at $T
\sim 13$. After this interaction, the eccentricity of the binary
reached $0.991$, resulting in the quick orbital decay by GW emission.

In this case, the high eccentricity is realized only after the last
strong interaction. However, there are several other periods when the
eccentricity was large, like around $T=7$ ($e \sim 0.97$) and $T=9$
($e\sim 0.94$). High values of   eccentricity are not very unusual after
strong interaction events.

\begin{figure}
\epsscale{0.75}
\plotone{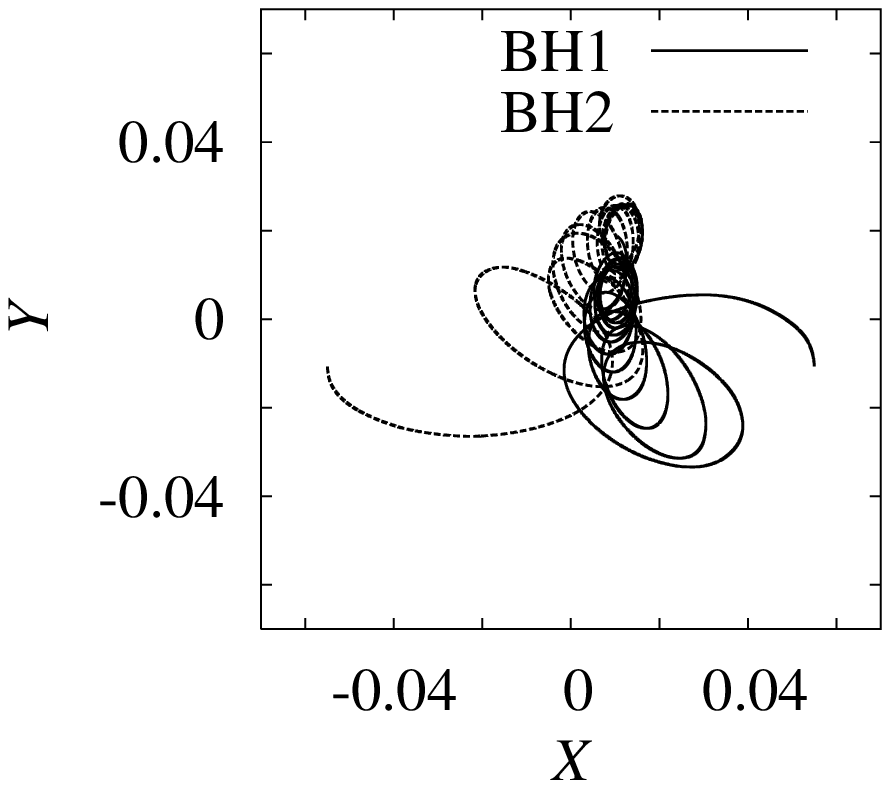}
\plotone{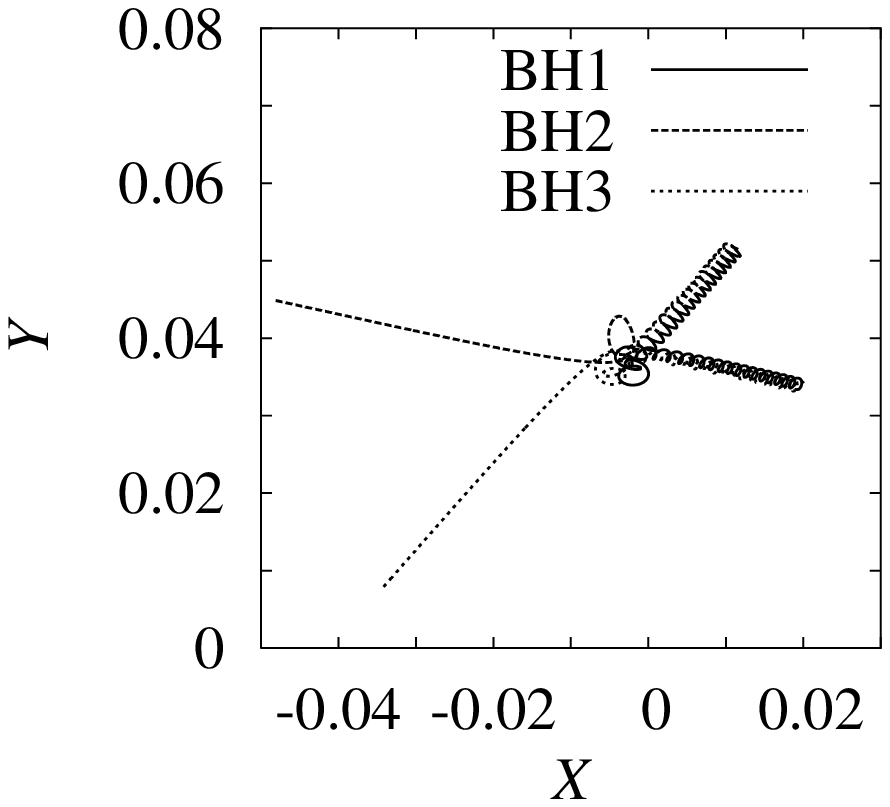}
\caption{Trajectories of the BHs before three body
encounter($T=0-1.8$) and just before and after the last strong
encounter ($T=12.85-12.95$). \label{fig3}}
\end{figure}

In figure \ref{fig3}, orbits of three blackholes in run 64kW7A projected on
the $x-y$ plane are shown. The top panel shows the trajectory of two
BH particles which are initially in a binary, for the period of $0\le
T\le 1.8$. The third BH particle is out of the plotting region and thus
not shown. Initially, the orbits decay quickly. As a result the orbital
period becomes shorter and the change in one orbit becomes smaller. In
the meantime, the center of mass of the binary moves along the  $y$
axis, because of the gravitational force from the third BH particle
which was initially located at (0,1,0) and falling down to the center
of the galaxy.

The bottom panel of figure \ref{fig3} shows the last strong
encounter. The binary BH (of BHs numbered 1 and 3) comes in from the
right side, and the third BH (BH2) comes in from the left side. After
the complex encounter, a new binary of BHs 1 and 2, with the very
high eccentricity, flies out to upper-right corner and BH3 to the
lower-left corner.

\subsection{Merging Probability}
\label{sect:statistics}

The last column of table \ref{tbl2} shows the fraction of runs in
which merger occurred. The divisor is the total number of runs
performed for the model and dividend is the number of runs in which
merging occured.
For runs of model 64kW7 with velocity dispersion  $\sigma=300 {\rm
km/s}$, in three out 
of five runs the BH binary merged before single BH is kicked out of
the parent galaxy. In two other cases, a single BH was kicked out. For
runs of model 64kW7High ($\sigma=600 {\rm km/s}$), two out of three resulted in
merging. All of runs of model 64kW7Low ($\sigma=150 {\rm km/s}$) resulted in ejection.

The total potential depth of typical giant ellipticals is somewhere
between galaxy models 64kW7 and 64kW7High. Thus, we can conclude that
a triple BH system, if formed in giant ellipticals, is likely to end
up in merging of two of the three BHs, leaving out a binary BH system.
THe Ejection of one BH is not impossible, though. In the remaining part of
this section, we check the effect of artificial parameters such as the
softening and the number of particles.

\subsection{Effect of the softening}

\begin{figure}
\plotone{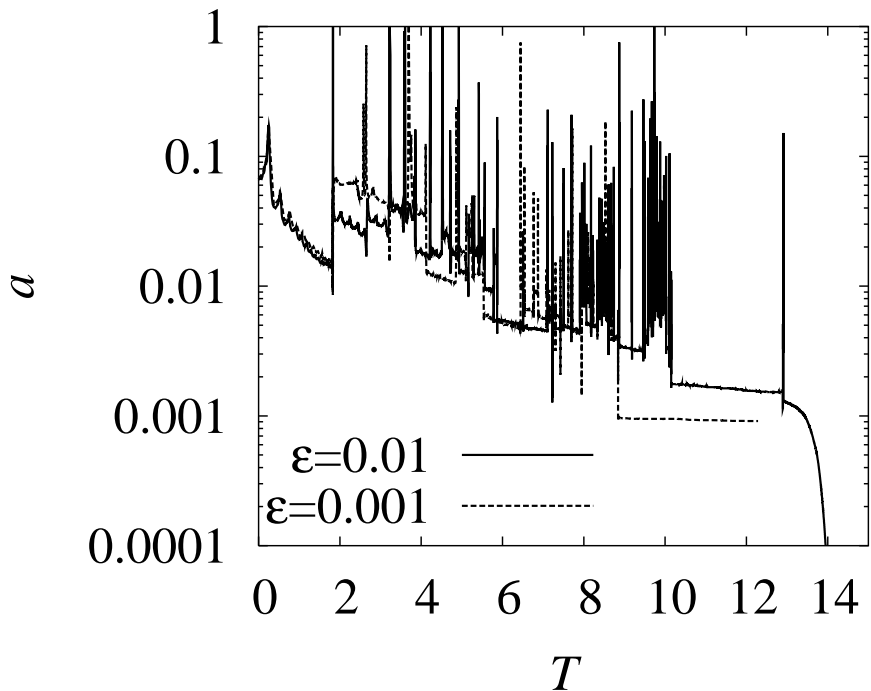}
\caption{
The evolution of semi-major axis for runs with different values of
the softening parameter $\epsilon_{ff}$ for field particles.  Solid 
and dashed curves are the results of runs with $\epsilon_{ff} =0.01$ and 0.001,
respectively. All other parameters are the same as those in model 64kW7.
\label{fig:compareEps}}
\end{figure}

In figure \ref{fig:compareEps}, we show the evolution of semi-major
axis of binaries of two runs with different softening parameters between
FS-FS. Solid and dashed curves correspond to runs with 
$\epsilon_{ff}=0.01$ and
$0.001$, respectively. Though the details are different, the overall
evolution is similar. We performed five runs with $\epsilon_{ff}=0.001$, and one
runs ended up in merging. This result is consistent with that for
$\epsilon_{ff}=0.01$.

\subsection{Effect of the number of particles}

\begin{figure}
\plotone{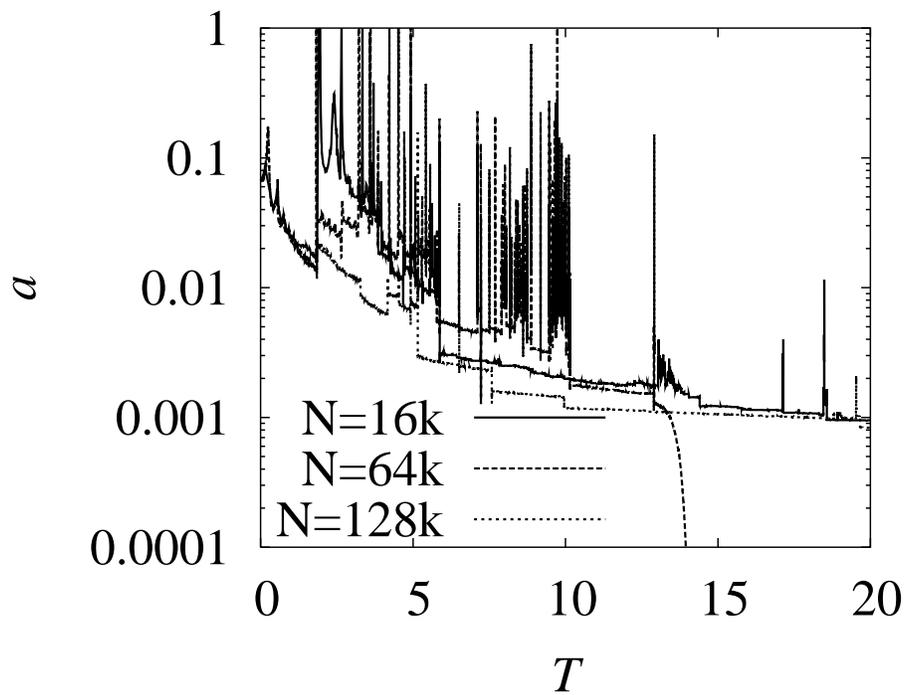}
\caption{
The evolution of semi-major axis for different values of $N$.
Solid, long-dashed and short-dashed curves are
the results of runs with  $N =16{\rm k}$,
64k and 128k, respectively.
All other parameters are the same as those in model 64kW7.
\label{fig:compareN}}
\end{figure}

In figure \ref{fig:compareN}, we show the evolution of semi-major
axis of binaries in runs with different number of particles for the
parent galaxy. Again, the overall behavior is not much different.
For $N=16{\rm k}$ and $N=128{\rm k}$, we again performed ten and five runs,
and four runs each ended up in merging. We can conclude $N$ has little
effect on the result. 

This result is quite different from that for the evolution of two
massive BHs (one BH binary) in the central region of the galaxy, where the
number of particles, {\it i.e.}, the relaxation timescale, determines
the evolution timescale of the BH binary.
In the case of two BHs, the loss cone is soon depleted, and the only way
for the field particles to interact with the BH binary is to enter
the loss cone through two-body relaxation. Thus, it is
quite natural that the evolution timescale depends on $N$. However,
in the case of a triple BH system, three BHs alone determines the
evolution, and field particles works just as the background potential
and the source for the dynamical friction. Thus, there is nothing like
loss-cone depletion. 

\subsection{Effect of the initial galaxy model}

In addition to models with $W_0=7$, we tried three runs with
$W_0=11$. As can be seen in table \ref{tbl2}, there is not much
difference in the final outcome. This result looks a bit
counter-intuitive, since, as we have seen in section
\ref{sect:statistics}, the merging probability does depend on the
depth of the potential relative to the speed of the light.

The reason why the outcome is not much different in cases with $W_0=7$
and $11$ is that initial binary BH hardens much more quickly in models
with $W_0=11$ than in models with $W_0=7$.
As a result, ejection of BH 
occurs after smaller number of triple interactions and the chance to 
reach high eccentricity is smaller for higher central density.
This difference more than compensates the deeper potential.

\begin{deluxetable}{ccccccc}
 \tabletypesize{\scriptsize}
 \tablecolumns{6}
 \tablecaption{Models for section 3\label{tbl2}}
 \tablewidth{0pt}
 \tablehead{
 \colhead{Model name} & \colhead{$N_f$}
& \colhead{$W_0$}
& \colhead{$\epsilon_{ff}$}
& \colhead{$\sigma$}
& \colhead{$c$} & \colhead{merging probability}
 }
\startdata
16kW7 &16k &7 &0.01 &300km/s &706 &$4/10$ \\
32kW7 &32k &7 &0.01 &300km/s &706 &$2/5$ \\
64kW7 &64k &7 &0.01 &300km/s &706 &3/5 \\
128kW7 &128k &7 &0.01 &300km/s &706 &4/5 \\
64kW7High &64k &7 &0.01 &600km/s &353 &2/3 \\
64kW7Low &64k &7 &0.01 &150km/s &1412 &0/3 \\
64kW11 &64k &11 &0.001 &300km/s &677 &1/3 \\
64kW7Eps &64k &7 &0.001 &300km/s &706 &1/5 \\
\enddata
\end{deluxetable}

\section{Result II --- General configurations}
\label{sect:general}

In section \ref{sect:planar}, we considered the simplest case in which
a binary is at the center of the galaxy and the third BH particle
falls from around the half-mass radius. This is clearly a rather
special case, and the result may be biased. In this section we
consider several different configurations.

\begin{figure}
\epsscale{0.49}
\plotone{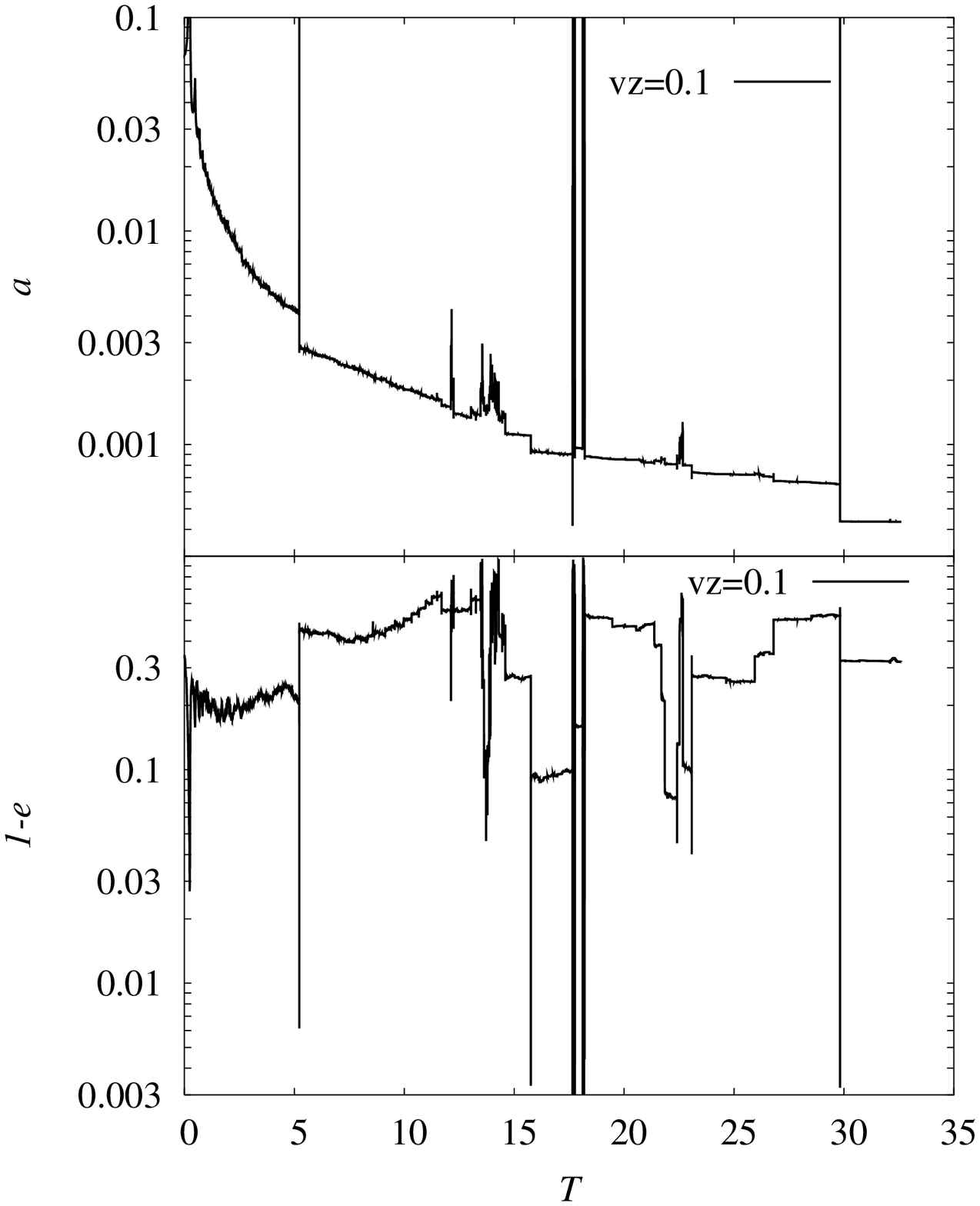}
\plotone{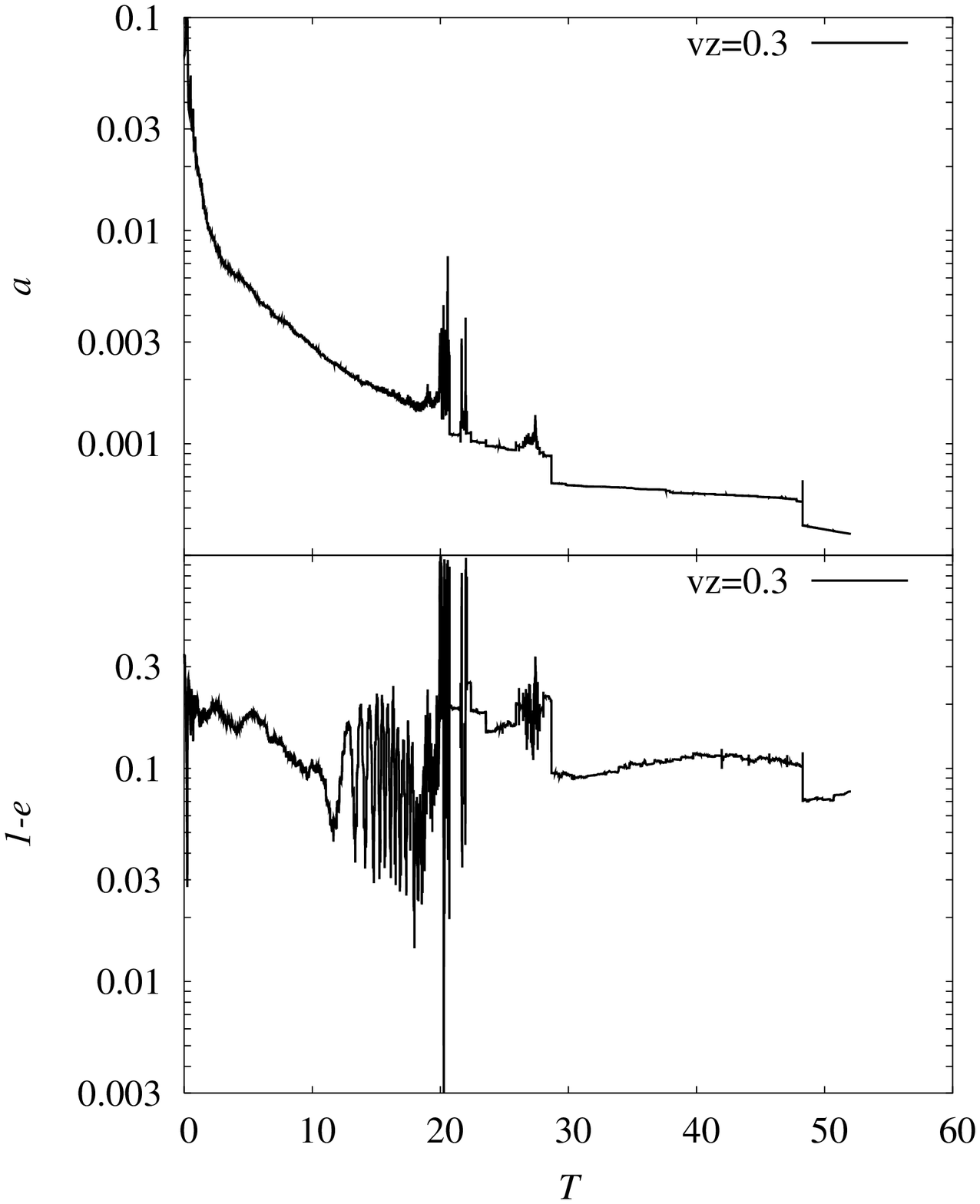}
\plotone{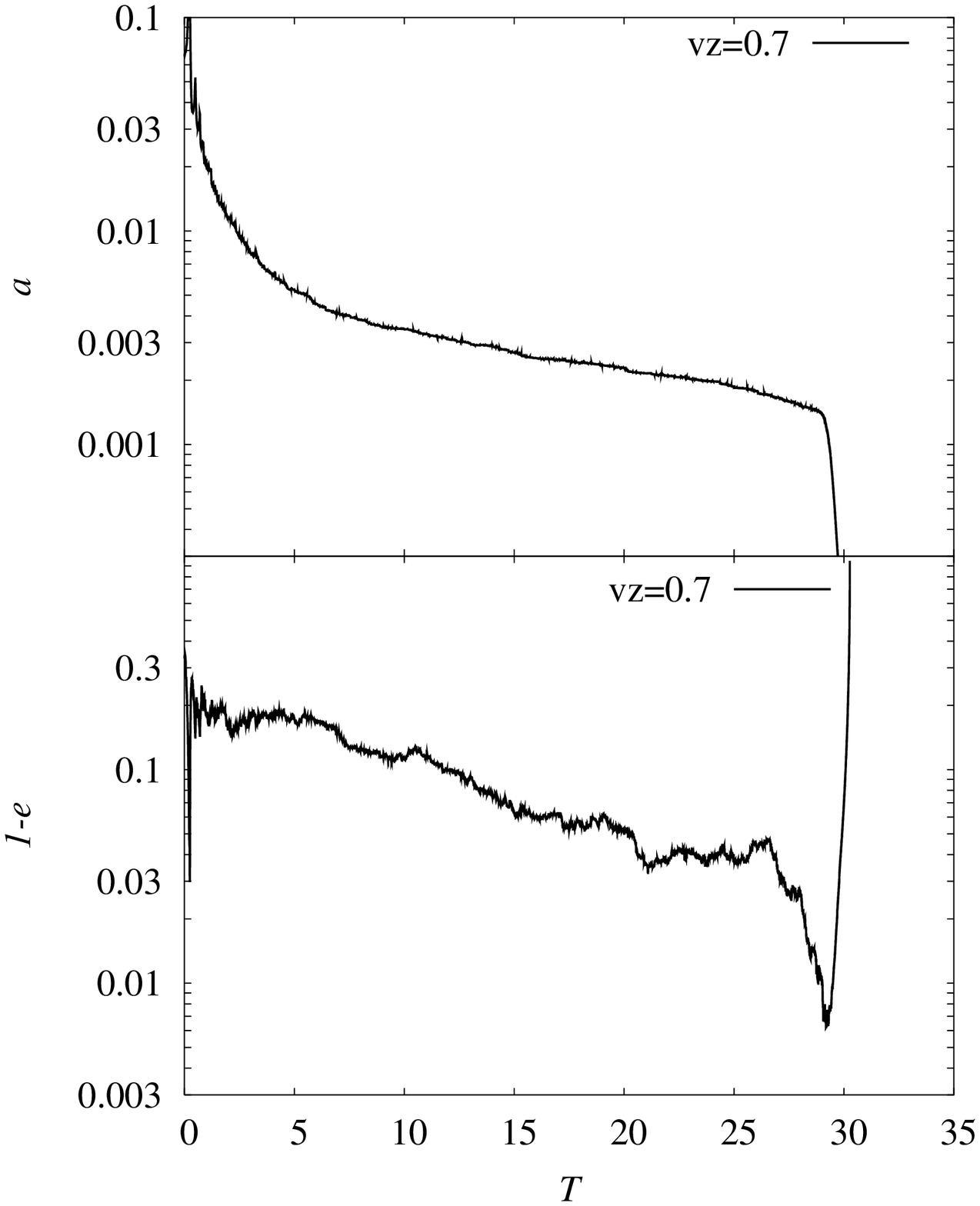}
\caption{
The evolution of the semi-major axis $a$ and eccentricity
 $e$ for runs
with non-zero initial $V_z$ of the third BH particle. Left top, right top, and
bottom panels show the results for $V_z=0.1,0.3,0.7$, respectively.
}
\label{fig6}
\end{figure}

Figure \ref{fig6} shows the result of runs in which the third BH has
the initial velocity perpendicular to the orbital plane of the binary
BH (runs 64kI90Vx, see Table \ref{tbl3}). The case with $v_z=0.1$ is qualitatively similar
to the result of free-fall runs (figure \ref{fig2}). Orbital elements
of the binary shows noncontinuous changes after strong encounters with
the third BH. The time interval for the encounter seems to be longer,
due to the larger total angular momentum of the triple BH system.

The case with $v_z=0.3$ shows quite different behavior. The
eccentricity shows periodic-like change for $T=11\sim18$, but the
semi-major axis does not show much change during the same period. This
behavior is driven by the Kozai mechanism. The period and magnitude
of the oscillation change in time, because the orbital elements of
both the inner and outer binaries changes due to the interaction with
other field stars.
Around $T=18$, the hierarchical system of three BHs became
unstable, and strong three-body interactions followed. 

In the case of $v_z=0.7$, the inner binary merged in the first cycle
of the Kozai cycles. 

\begin{figure}
\epsscale{0.49}
\plotone{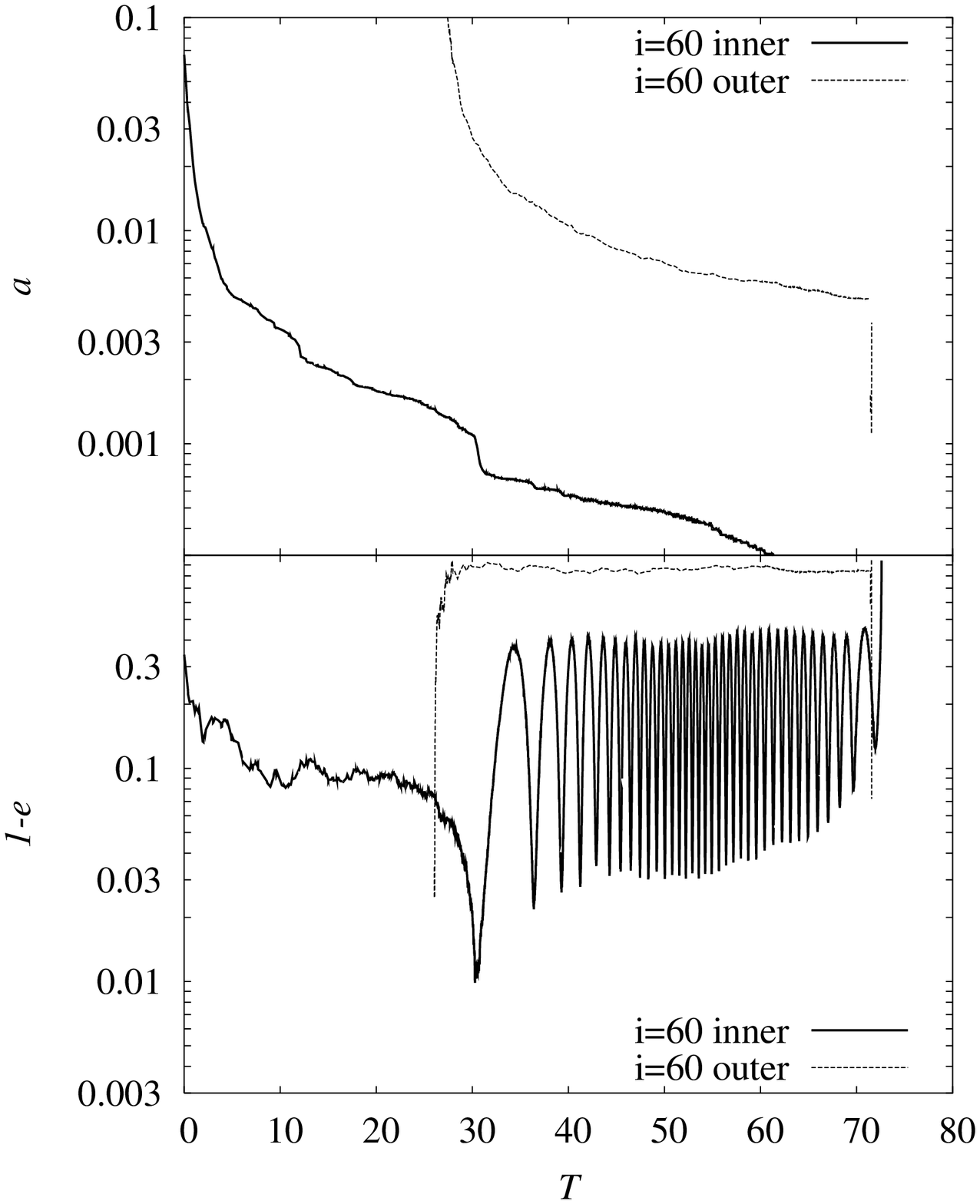}
\plotone{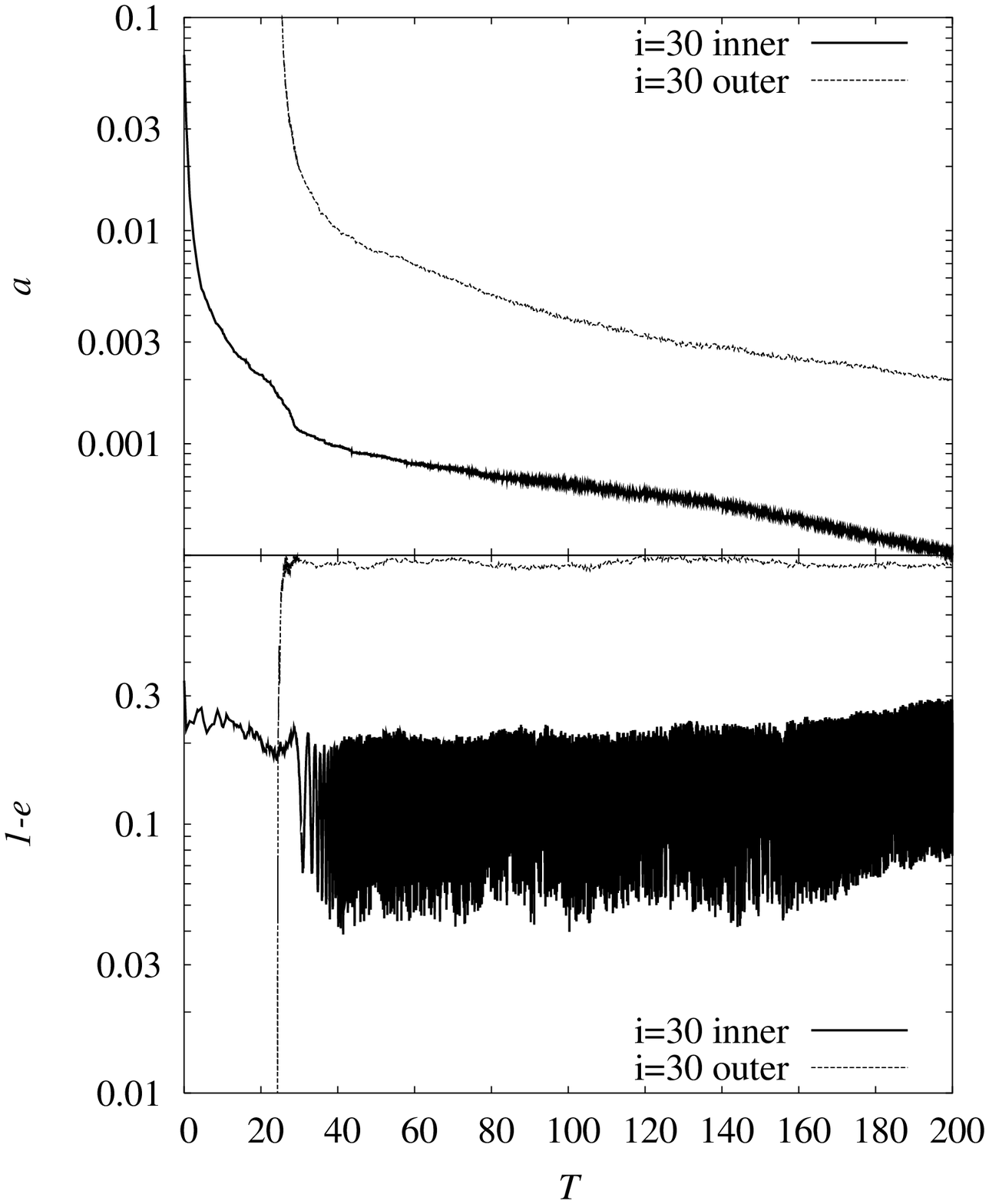}
\plotone{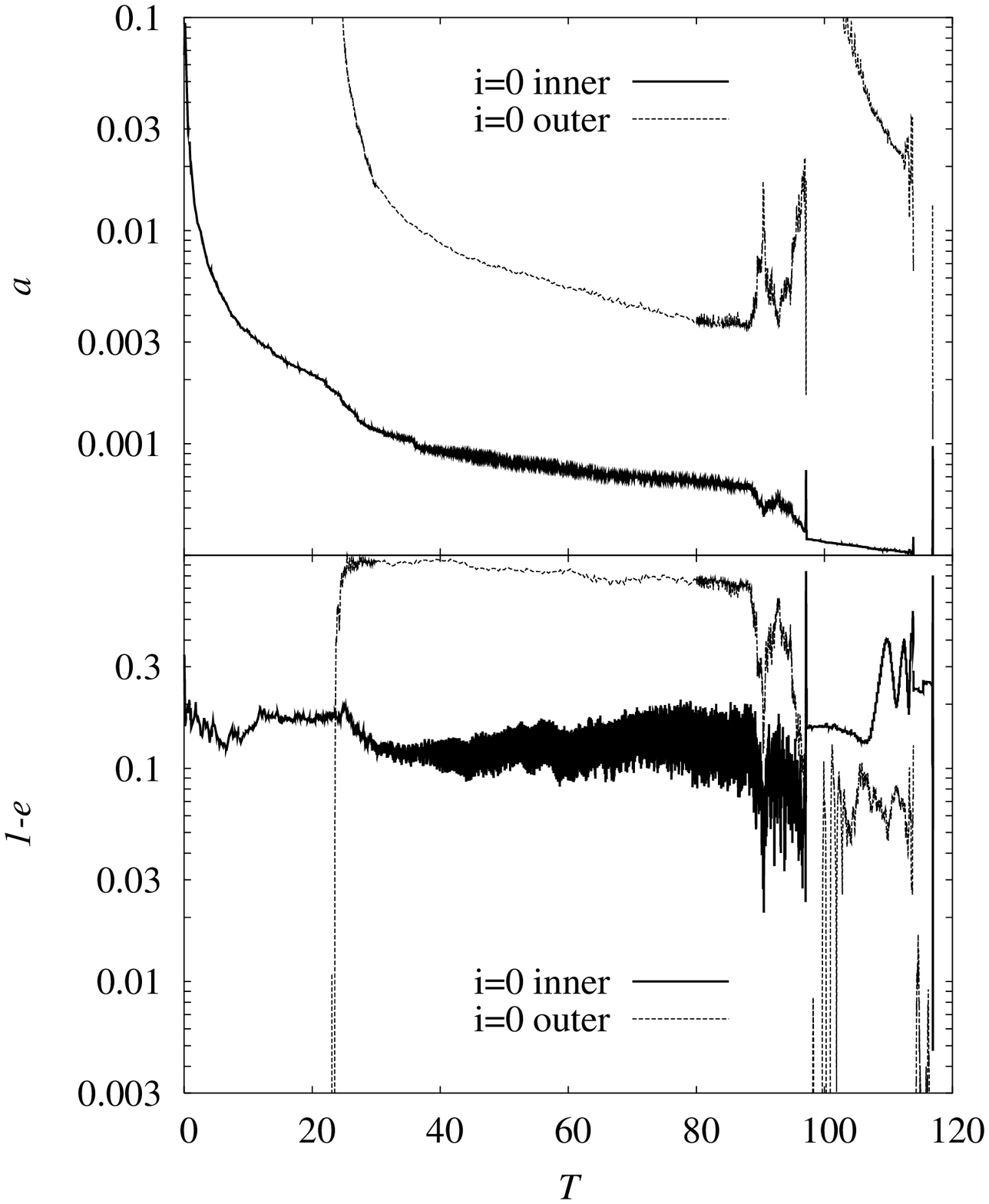}
\caption{
The evolution of the semi-major axis $a$ and eccentricity $e$ for runs
with initial $|V|=0.7$ for the third BH particle. Initial velocity of
the third BH is in $yz$ plane and the angle with the $xy$ plane is 60
 (left top),
30 (right top) and 0 (bottom) degree.
Each panel shows the
semi-major axis and the eccentricity  of inner (thick solid line) and 
outer binaries (thin dashed line).
\label{fig7}}
\end{figure}

Figure \ref{fig7} shows the results of three runs with different
initial inclination of the orbital plane of the third BH.
As we've seen in figure \ref{fig6}, in the extreme case of large $v_z$
perpendicular to the binary orbital plane, the end result is the
merging in the first Kozai cycle. In the case of $i=\pi/3$, Kozai
cycle drives oscillations of large amplitude, and semi-major axis
shrinks each time when the eccentricity reached the
peak.
In the case of $i=\pi/6$, BHs neither merge nor escape until $T=200$.
Kozai cycles are also observed.
The maximum eccentricity in one cycle is, however,  not high enough to
drive the merging of inner binary.
In the case of the coplanar orbits, Kozai cycle of a small magnitude is
observed for $T=20$ to 90. Since the initial orbit is coplanar, there
cannot be any Kozai cycle if three BHs remain exactly in the same
orbital plane. However, since the number of particles is finite,
fluctuation in the forces from field particles introduces random
changes in the inclination of inner and outer binaries, resulting in
the configuration for which the Kozai cycle can show up.

To summarize the results of 3D triple cases, a hierarchical triple BH
system is often formed when the third BH falls in.  In such a triple
system, a stable oscillation in eccentricity (and inclination) driven
by Kozai mechanism is
often observed. In a fair fraction of the cases, this oscillation
results in the merging of the inner binary by GW emission.

\begin{deluxetable}{rccccccl}
 \tabletypesize{\scriptsize}
 \tablecolumns{6}
 \tablecaption{Models for section 4\label{tbl3}}
 \tablewidth{0pt}
 \tablehead{
 \colhead{Model name} & \colhead{$N_f$} 
& \colhead{$W_0$}
& \colhead{$i$}
& \colhead{$|V|$}
& \colhead{$\sigma$}
& \colhead{$c$} & \colhead{final state}
 }
\startdata
64kI90V1 &64k &7 &90 &0.1 &300km/s &706 &escape \\
64kI90V3 &64k &7 &90 &0.3 &300km/s &706 &escape \\
64kI90V7 &64k &7 &90 &0.7 &300km/s &703 &merge \\
64kI60V7 &64k &7 &60 &0.7 &300km/s &703 &merge \\
64kI30V7 &64k &7 &30 &0.7 &300km/s &703 &neither merge nor escape by $T=200$\\
64kI0V7 &64k &7 &0 &0.7 &300km/s &703 &merge \\
\enddata
\end{deluxetable}

\section{Discussion}

\subsection{Escaping cases and free-floating BHs}

We have seen that the merging of the BH binary is more likely than the
ejection of single BH (or also the binary BH). However, the ejection
is not a rare event. Out of the five runs with $N=64k$ and
$\sigma=300{\rm km/s}$  listed in table \ref{tbl2}, two runs ended
up in the ejection. Though this result is somewhat biased by the
shallow galactic potential, runs with deeper potential still gives a
fair number of escaping events.

Thus, our result suggests that, though in most cases the binary BH do
merge, there may be cases in which one or more BHs are kicked out. In
other words, there may be a fair number of massive BHs
free-floating  in intergalactic field.

An interesting question is if such a floating BH can be observed. One
possibility is through strong lensing, but since the Einstein radius,
when placed around $z=0.5$, is around $10^{-2}$ arcsec,
it would be hard to detect it even with HST. Another possibility
is the detection of X-ray emission  by Bondi accretion of intracluster
hot gas, if BH is in a cluster of galaxy. The luminosity $L$ is given
by 
\begin{equation}
L = \eta \dot{m}_B c^2 ,
\end{equation}
where $\eta$ is the radiative efficiency, $\dot{m}_B$ is the accretion
rate and $c$ is the light velocity. The accretion rate is given by 
\begin{equation}
\dot{m}_B = F(\gamma) \pi G^2 m_B^2 \rho_{\infty} / a_{\infty}^3 ,
\end{equation}
where $F(\gamma)$ is a function of adiabatic index of the gas
$\gamma$, 
$\rho_{\infty}$ and $a_{\infty}$ are density and sound speed at
infinity, respectively \citep{bondi52}.
We take $\eta =0.1 $ and $F(\gamma)=1$.
For a  BH of  $10^8 M_{\odot}$ in the Virgo cluster, 
whose typical hot gas density and sound speed is given by
\begin{equation}
 \rho_{\infty} \sim  1.7\times 10^{-28}
 \frac{n}{1.0\times 10^{-4}\rm{cm^{-3}}}\rm{g/cm^{3}} ,
\end{equation}
\begin{equation}
a_{\infty} \sim  4.3 \times
 10^{7}\Bigl(\frac{kT}{2\rm{keV}}\Bigr)^{1/2}\rm{cm/s} , 
\end{equation}
where $n$ and $kT$ are the number density and temperature
of the intracluster hot gas and 
we adopted $n=1.0\times 10^{-4}\rm{cm^{-3}}$ 
and $kT=2\rm{keV}$ \citep{Kikuchi00} as typical values,
thus the luminosity would be
\begin{eqnarray}
L_{virgo} &=& 9.0 \times 10^{19} \dot{m}_B \rm{erg/s} \\
 &=& 7.9 \times 10^{37} \Bigl(\frac{m_B}{10^8M_{\odot}}\Bigr)^2\Bigl(
  \frac{n}{1.0\times 10^{-4}\rm{cm^{-3}}}
  \Bigr)\Bigl(\frac{kT}{2\rm{keV}}\Bigr)^{-3/2} .
\end{eqnarray}
Thus, it can be observed as a moderately luminosity X-ray source.
The question is how it can be discriminated from AGNs.

\subsection{Conclusion}

In this paper, we investigated the dynamical evolution of triple black
hole systems in  the central region  of a galaxy. We found that in most of cases
two black holes merge through GW radiation. The merging timescale by
GW radiation is greatly reduced by high eccentricity of the binary
BH. The high eccentricity is realized by two different
mechanisms. First one is the random change of the eccentricity after
binary-single BH interaction as suggested by \citet{ME94}, and the
other is the Kozai mechanism\citep{kozaimec1}. Even so, there may be a
fair number of free-floating BHs ejected out of the parent galaxy by
a slingshot. They might be observed as X-ray sources, if they stay
within a cluster of galaxies.

\acknowledgments
We thank Tatsushi Matsubayashi and Toshiyuki Fukushige for stimulating discussions and useful 
comments.
We are also grateful to Keigo Nitadori for preparing the GRAPE-6A library.

This research is partially supported by the Special Coordination Fund
for Promoting Science and Technology (GRAPE-DR project), Ministry of
Education, Culture, Sports, Science and Technology, Japan.

\appendix

\end{document}